\documentclass[doublecol]{epl2} 
\usepackage{graphics}
\usepackage[utf8]{inputenc}
\usepackage[pageanchor,hyperindex]{hyperref}
\usepackage{amsmath}
\usepackage{amsfonts}
\usepackage{graphics}

\title{On the trajectories and performance of Infotaxis, \\
an information-based greedy search algorithm}
\shorttitle{Search trajectories and performances of Infotaxis}
\author{C. Barbieri\inst{2}, S. Cocco\inst{1,2}, R. Monasson\inst{1,3}}
\shortauthor{C. Barbieri, S. Cocco, R. Monasson}

\institute{                    
\inst{1} Simons Center for Systems Biology, Institute for Advanced Study, Einstein Drive, Princeton, NJ 08540, USA \\
\inst{2} Lab. Physique Statistique de l'ENS, CNRS and Univ. Paris 6 - 24 rue Lhomond, 75231 Paris, France\\ 
\inst{3} Lab. Physique Th\'eorique de l'ENS, CNRS and Univ. Paris 6 - 24 rue Lhomond, 75231 Paris, France}

\pacs{05.40.-a}{Fluctuation phenomena, random processes, noise, and Brownian motion} 
\pacs{02.50.Tt}{Inference Methods}
\pacs{87.19.lt}{Sensory systems: visual, auditory, tactile, taste, and olfaction}

\abstract{We present a continuous-space version of Infotaxis, a search algorithm where a searcher greedily moves to maximize the gain in information about the position of the target to be found. Using a combination of analytical and numerical tools we study the nature of the trajectories in two and three dimensions. The probability that the search is successful and the running time of the search are estimated. A possible extension to non-greedy search is suggested. }

\begin{document}
\maketitle

\section{Introduction}

Reaching a target with limited information is a fundamental task for living organisms. Small organisms, such as bacteria and eukaryotic cells, are thought to estimate and ascend the gradient of nutrient concentration, a process called chemotaxis \cite{ADLE66,BERG75}.  At the scale of a larger organism the Reynolds number is higher\cite{BALK02}: most biologically relevant chemical fields become turbulent and dilute. As a result, the trajectories of insects following odor traces appear much more complex than those of smaller organisms \cite{VICK94}.  The modeling of search processes in presence of noisy information is important not only for biology, but also for robotics \cite{LOCH09,MORA10}.

Assume the search has proceeded for some time, and the searcher has received some hits, {\em i.e.} has detected some molecule of odor sent by the target, along the trajectory.  How should the searcher move next? The timing and locations of the hits, as well as the absence of hits along the remaining parts of the trajectory all provide useful information about the location ${\bf y}$ of the target. In Bayesian terms, this defines a posterior probability $P_t({\bf y})$ for the position of the target. Going towards the maximum of $P_t$ is not an optimal strategy as the maximum does generally not coincide with the target, especially in the initial stage of the search process. Recently, Vergassola, Villermaux and Shraiman proposed an alternative strategy, called Infotaxis \cite{VERG07}. The searcher moves to maximize the (expected) gain in information about the location of the target, that is, the loss in the entropy of the distribution $P_t({\bf y})$. As the search goes on, the entropy typically decreases, until the source is finally located. The strategy naturally balances the needs for exploration (harvesting more information about the target location) and exploitation (going towards the maximum of $P_t$). Infotaxis was implemented and tested on two-dimensional square lattices\footnote{Few short trajectories were obtained on small three-dimensional lattices in \cite{MASS09}.}: the target was almost always found, and the distribution of search time appeared to decay exponentially.

Yet some important questions about Infotaxis remain open. First, how well does the algorithm perform in three dimensions? In addition to its practical interest, this question arises naturally in the context of Brownian motion theory. Purely random walks are space-filling in two dimensions, and transient in higher-dimensional spaces. Finding a target in three dimensions is therefore much harder, and constitutes a real test for the capabilities of Infotaxis.  Secondly, how dependent of the underlying lattice are the results reported in \cite{VERG07}? Realistic descriptions of animal behavior or implementations in biomimetic robots require us to consider continuous spaces. In addition the presence of a lattice introduces anisotropies, while the odor propagation model used in \cite{VERG07} was isotropic. Thirdly, two-dimensional trajectories seem to exhibit spiral-like shapes. How precisely can we characterize those spirals, and what are their counterparts in three dimensions?

In this letter, we derive the equation of motion for the Infotaxis searcher in the  continuous space. We then introduce an algorithm to solve this equation \footnote{The code is publicly available from \url{http://www.lps.ens.fr/~barbieri}.}. The performances of Infotaxis, {\em i.e.} the probability of success and the distribution of the search times are studied in $D=2$ and 3 dimensions. The spiral- and coil-like shape of the search trajectories for, respectively, $D=2$ and 3, and the pinning of the trajectory around the received hits are investigated analytically and numerically. We show that the motion of a searcher receiving an average and deterministic signal is a good predictor of the typical properties of the motion in the presence of stochastic hits. Finally we discuss a possible extension to a non-greedy search strategy, which could help reduce pinning effects.

\section{Equation of motion for the searcher}

A point-source in ${\bf y}^*$ emits particles, which diffuse in space, and have a finite lifetime. In the stationary regime, the probability per unit of time to encounter a particle in ${\bf x}$ is denoted by $R({\bf y}^*-{\bf x})$ \cite{VERG07}[Supplementary information]. Function $R$ has an integrable divergence at the origin ($R(u)\sim -\log u$ in $D=2$, $\sim \frac 1u$ in $D=3$ dimensions, when $u\to 0$), and exponentially decreasing tails for large distances $u$. In the following distances are measured in the unit of the decay length of $R$. The unit of time is the inverse of the rate of emission of particles by the source, divided by the (dimensionless) linear size, $a$, of the searcher for $D=3$, or multiplied by $\log(1/a)$ for $D=2$ \cite{VERG07}. 

Let ${\bf x}(t)$ be the position of the searcher at time $t$. We denote by $N_H$ the number of particles detected (called hits) at earlier times, $0\le t_i\le t$, with $i=1,\ldots , N_H$. Based on those hits, the searcher can draw a probabilistic map over the possible locations ${\bf y}$ of the source.  In the Bayesian framework, the posterior probability density $P_t({\bf y})$ for the location of the source is the (normalized) product of the probabilities of having detected the $N_H$ particles at locations ${\bf x}(t_i)$, times the probability of not having detected any particle at other locations along the trajectory, times the prior probability density $P_0$ over ${\bf y}$,
\begin{equation}\label{eq:py}
P_t({\bf y})\propto \prod_{i=1}^{N_H} R({\bf y}-{\bf x}(t_i)) \; e^{-\int_0^t \upd t^\prime R({\bf y}-{\bf x}(t^\prime))}\; P_0({\bf y})\ .
\end{equation}
Hence, $P_t$ diverges where the hits have been received and vanishes in the other places along the trajectory. In the following we will use brackets to denote averages over this posterior distribution,
\begin{equation}\label{bracket}
\langle f ({\bf y})\rangle _{{\bf y};t} = \int \upd{\bf y} \; P_t({\bf y}) \; f({\bf y})\ .
\end{equation}
Assume now that the searcher stays in ${\bf x}(t)$ during an infinitesimal time $\delta t$.  The number of hits, $n$, received during this time interval is a stochastic variable equal to zero or one, with probabilities $p(0|{\bf y}) = 1-\delta t\, R({\bf y}-{\bf x})$ and $p(1|{\bf y}) = \delta t\, R({\bf y}-{\bf x})$, depending on
the location ${\bf y}$ of the source. In the language of information theory, the particle emission and detection system can be thought as a noisy channel, and $n$ is the output message associated to the input codeword ${\bf y}$. The mutual information $\delta I$ between $n$ and ${\bf y}$ is 
\begin{equation}
\delta I=\!\!\!\!\sum_{n=0,1}\left< P(n|{\bf y}) \log\left(\frac{ P(n|{\bf y})}{ \langle P(n|{\bf y}') \rangle _{{\bf y}';t}}\right)\right>_{\!\!\!{\bf y};t} \!\!\!\!=-\delta t \;  V_t\big( {\bf x}(t)\big) 
\end{equation}
up to $O(\delta t^2)$, where
\begin{equation}V_t({\bf x})= \left< R({\bf y}-{\bf x})  \log \left( \frac{\langle R({\bf y}'-{\bf x})  \rangle _{{\bf y}';t}} { R({\bf y}-{\bf x})  }\right)\right>_{{\bf y};t} \ , \label{eq:pot}
\end{equation}
is the entropy rate of the posterior distribution $P_t$. Infotaxis stipulates that $\delta I$ should be maximized, or, equivalently that $V_t$ should be minimized, {\em i.e.} made as negative as possible. In other words, interpreting $V_t$ as a potential, the searcher should descend the gradient of $V_t$. A natural equation of motion is then
\begin{equation}\label{eq:eqsearcher}
\gamma(t)\; \dot {\bf x}(t)=-\nabla _{\bf x} V_t \big ({\bf x}(t)\big)\ ,
\end{equation}
where $\gamma(t)$ plays the role of a friction coefficient. A possible choice is $\gamma(t)=|\nabla _{\bf x} V_t \big ({\bf x}(t)\big)|/v_{0}$, to keep the modulus of the velocity fixed and equal to $v_{0}$. This is close to the lattice version of \cite{VERG07}, where the searcher could either stay immobile or move by one lattice site, and the velocity could take only one non-zero value. In general, $\gamma(t)$ can be any positive function of the time. In the following, we will first consider the case $\gamma(t)=\gamma$ constant. We will later discuss to what extent the trajectories and the performances change when $\gamma$ varies with the time.

\section{Numerical integration}

The equation of motion (\ref{eq:eqsearcher}) is highly non-linear and depends on the whole history of the search process through the posterior probability (\ref{eq:py}). To solve equation (\ref{eq:eqsearcher}) numerically  we discretize the time with a step $\Delta t$. The positions ${\bf x}_\ell$ of the searcher at the instants $t_\ell = \ell \; \Delta t$, where $\ell$ is a positive integer, are memorized, as well as the occurrences (times) of the hits. The amplitude of the $\ell^{th}$ elementary move is estimated from (\ref{eq:eqsearcher}): ${\bf x}_{\ell+1}-{\bf x}_\ell = -\frac{\Delta t}{\gamma (t_{\ell})}\; \nabla _{\bf x} V_t \big ({\bf x}_\ell\big)$. The calculation of the gradient of the potential requires to estimate averages over the space ${\bf y}$  with measure $P_t$ (\ref{bracket}). To do so, we use the importance sampling Monte Carlo method \cite{HAMM64}. The term inside the bracket in (\ref{eq:pot}), and its gradient with respect to ${\bf x}$ are exponentially decreasing functions of the radial distance $u=|{\bf x}_\ell -{\bf y}|$. We thus perform the change of variable $u = u_0 (1-v)/v$, where $v\in ]0;1]$, and $u_0$ is a scale parameter. We draw $N_{MC}$ values of the new variable, $v_a$, $a=1, \ldots, N_{MC}$, uniformly and at random, and calculate the corresponding $u_a$. Angular variables ${\bf \Omega} _a$ are uniformly sampled on the unit sphere or circle to obtain the points ${\bf y}_a= {\bf x}_\ell + u_a \, {\bf \Omega} _a$ in the original space. The  corresponding probabilities $P_t({\bf y}_a)$ are computed using Simpson's method to calculate the integral over time in (\ref{eq:py}).

The algorithm of \cite{VERG07} stored and updated $P_{t}$, which made the computational time linear in $t$ and in the size of the lattice. Our procedure recalculates $P_t$  at each time step, which makes the computational time quadratic in $t$, and independent of the {\em a priori} infinite size of the space. Errors on $P_t$ do not accumulate with time, and the accuracy is directly controlled by the number of Monte Carlo sampling points, $N_{MC}$. In addition, the map $P_t$ is guaranteed to be accurately determined where it really matters, {\em i.e.} in the vicinity of searcher.

%%%%%% beginning of search process

\section{Initial stage of the search}

\begin{figure}
\begin{center}
\includegraphics[width=8cm]{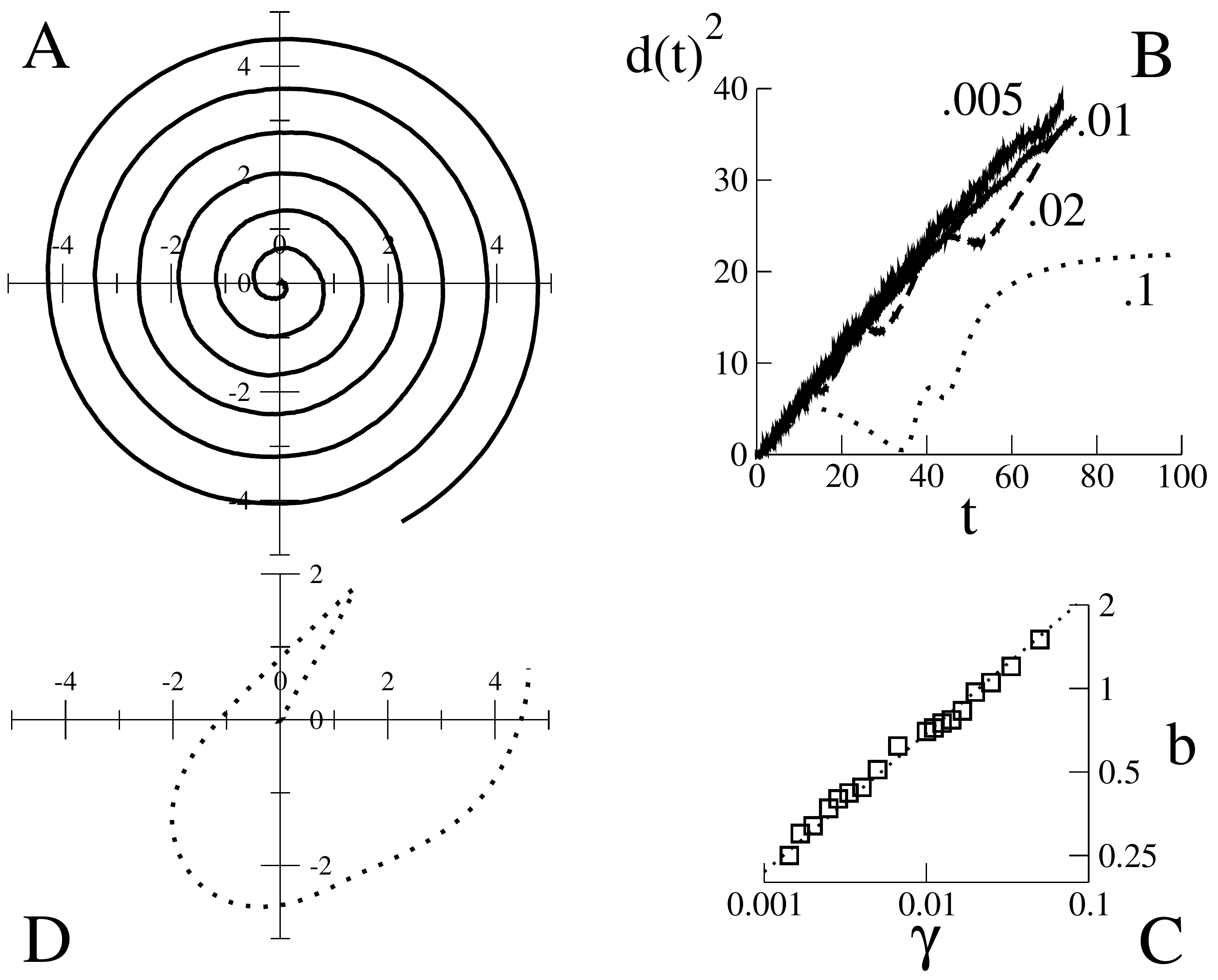}
\caption{Two-dimensional trajectories for $\gamma=.01$ ({\bf A}) and $\gamma=.1$ ({\bf D}), after one hit is received at time $t=0$ $(P_0=R$).  {\bf B.}~squared distance to the origin, $d(t)^2$, vs. time $t$ for four values of $\gamma$. {\bf C.} spacing $b$ between turnings vs. $\gamma$. The dotted line has slope $\frac 12$. Integration is done with $N_{MC}=10^4$ points.}
\label{fig:spiral2D}
\end{center}
\end{figure}

We first focus on the initial stage of the search process, which strongly depends on the \emph{a priori} distribution, $P_0$.  A possible choice is the {\em one-hit} prior, $P_0=R$, which means that the search starts only when a first hit is received at time $t=0$ \cite{VERG07}.  In two dimensions and before subsequent hits are detected, the search trajectories are Archimedean spirals \cite{tesi,MASS09} for a large range of values of $\gamma$ (Fig.~\ref{fig:spiral2D}A). The squared distance of the searcher to the origin at time $t$, $d(t)^2$, increases linearly with $t$, and is, to a large extent, independent of $\gamma$ (Fig.~\ref{fig:spiral2D}B). The spacing $b$ between successive turnings is independent of time and increases as $\sqrt \gamma$ (Fig.~\ref{fig:spiral2D}C); the velocity $|\dot {\bf x}| \sim \frac 1b$ decreases with the friction $\gamma$ as expected. 

The increase of $b$ can be intuitively understood: the larger $\gamma$, the longer the searcher spends along the trajectory without receiving hits, and the more likely is the source to be located far away. For values of $\gamma (> .08)$ such that $b$ would exceed the length ($=1$) over which $R$ decreases, the spiral nature of trajectories breaks down (Fig.~\ref{fig:spiral2D}D). We have run simulations with a modified potential $V_t$, where the arguments ${\bf y}-{\bf x}$ and ${\bf y}'-{\bf x}$ of the functions $R$ in (\ref{eq:pot}) were divided by a large factor ($=10$). The regular spirals then disappeared, and looked like the trajectory in Fig.~\ref{fig:spiral2D}D, even for small values of $\gamma$.

Simulations with other prior distributions show that the long distance behavior of $P_0$ is critical to the existence of spiral trajectories, while the behavior of $P_0$ close to the origin is irrelevant.  Choosing $P_0 ({\bf y}) \sim \exp( -|{\bf y}|/y_0)$ we obtain spirals as long as $y_0$ is not too large. The reason is that the spacing $b$ is proportional to $y_0$: spirals explore as much space as allowed by the prior. Spirals breakdown for the ($\gamma$-dependent) value of $y_0$ such that $b$ exceeds 1.  

\begin{figure}
\begin{center}
\includegraphics[width=8cm]{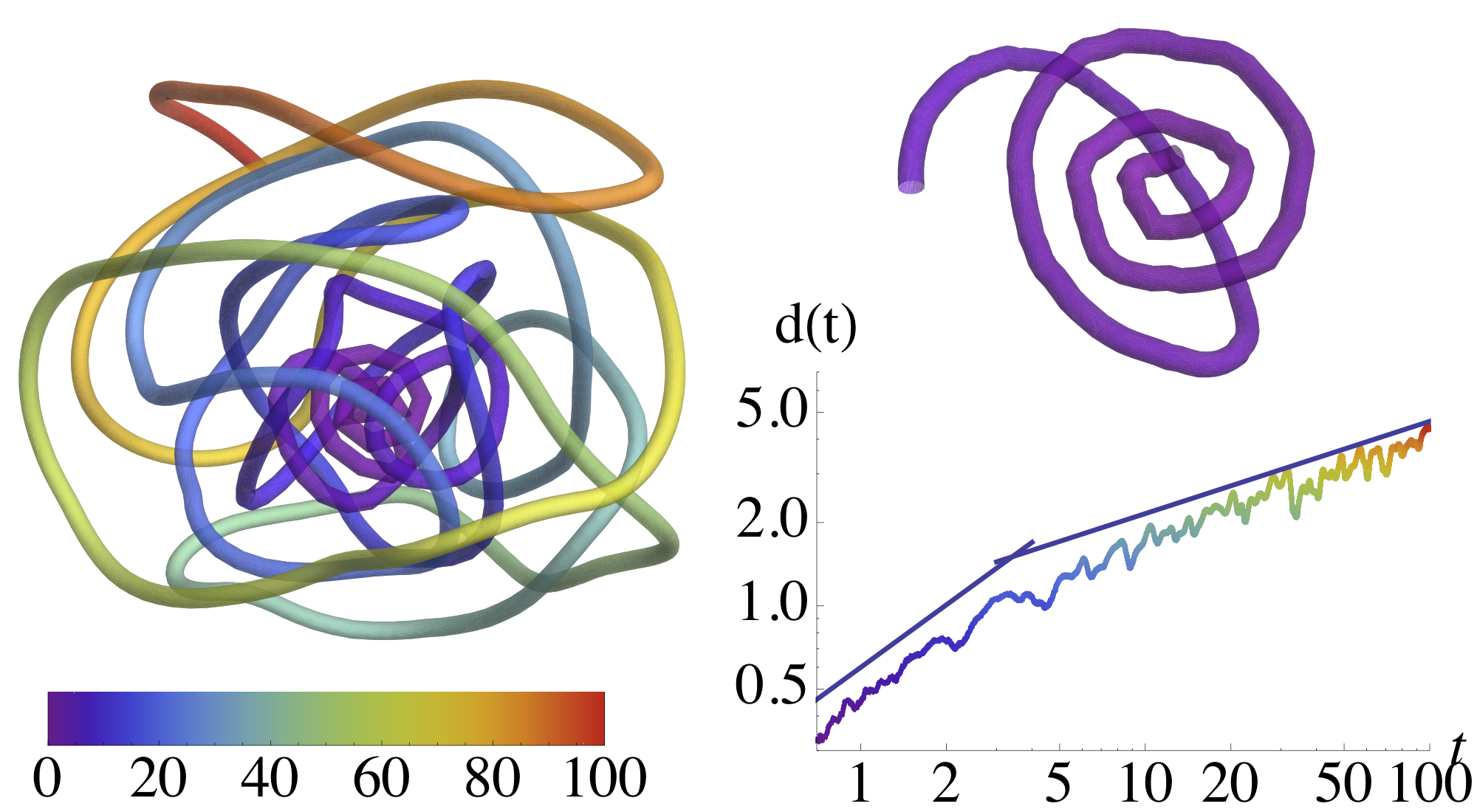}
\caption{A three-dimensional trajectory in the absence of hit for $\gamma=.01$ (left), with its quasi two-dimensional initial portion (top); the time axis is color coded. Bottom: distance to the origin, $d(t)$, compared to the power laws $t^{.75}$, then $t^{1/3}$.}\label{fig:spiral3D}
\end{center}
\end{figure}

Search trajectories in three dimensions display a more complex structure than their two-dimensional counterparts. Figure \ref{fig:spiral3D} shows the motion of the searcher with the {\em one-hit} prior, $P_0=R$. Roughly speaking, the trajectory is constituted of subsequent shells of increasing radii, which are densely covered before a new shell is built. The distance to the origin, $d(t)$, is compatible with $t^{1/3}$ at large times, but grows faster at smaller times (Fig.~\ref{fig:spiral3D}). To better understand how trajectories develop in three dimensions, we have resorted to a small--${\bf x}$ expansion of the potential $V_t$. The relevant contributions to the equation of motion are, up to cubic order,
\begin{eqnarray}\label{eq:eqapprox}
&\gamma &\!\!\! \dot {\bf x}(t) =\alpha _1(t) \; {\bf x}(t)+\alpha _2(t) \int_0^t \upd t^\prime \;{\bf x}(t^\prime) + \\
&&\int_0^t \!\!\! \upd t^\prime \;{\bf x}(t^\prime)  \bigg[ \beta _1(t) |{\bf x}(t^\prime)|^2 + \beta _2(t)\; {\bf x}(t^\prime) \cdot \int _0^t \!\!\! \upd t''{\bf x}(t'') \bigg]
\nonumber
\end{eqnarray} 
where the coefficients $\alpha _i(t)$ have explicit analytical expressions. When ${\bf x}$ is very small, only the linear terms matter. As $\alpha_1\simeq \frac {\sqrt 2}{3e}\frac{\log t}t>0$ for large $t$, the trajectory tends to follow a line radiating from the origin. However, the straight line is unstable against local bending since $\alpha_2\simeq -\frac{3\sqrt 3}{e^2}\frac{\log t}{t^2}<0$. The trajectory thus acquires a spiral shape confined within the plane spanned by ${\bf x}(0)$ and $\dot {\bf x}(0)$. The presence of cubic terms ($\beta_1,\beta_2>0$), which are not constrained to lie in this plane, eventually lead to a cross-over from the quasi-bidimensional spiral to a fully three-dimensional trajectory (Fig.~\ref{fig:spiral3D}). 

Replacing coefficients $\alpha_{1},\alpha_{2}$ with the smaller values $\alpha_{1}=\frac{a_{1}}t,\alpha_{2}=-\frac{a_{2}}{t^{2}}$, with $a_{1},a_{2}>0$, allows for an exact resolution of (\ref{eq:eqapprox}). A spiral is found when $a_1<a_2$, with a radius and an angle growing as, respectively, $t^\omega$ and  $\eta \log t$, where $\omega =\frac{a_{1}-\gamma}{2\gamma}$ and $\eta =\sqrt{4a_2\gamma-(\gamma+a_{1})^2}$. As $\gamma$ increases so does the angular velocity ($\propto \eta$), while the growth exponent $\omega$ diminishes: spirals stop growing if $\gamma$ is too large, a fact reminiscent of Fig.\ref{fig:spiral2D}D. Numerical resolution of (\ref{eq:eqapprox}) shows that this scenario is qualitatively unchanged when the logarithmic factors in $\alpha_{1},\alpha_{2}$ are taken into account.

%%%%%%% pinning

\section{Pinning after a hit}

\begin{figure}[h]
\includegraphics[width=8.6cm]{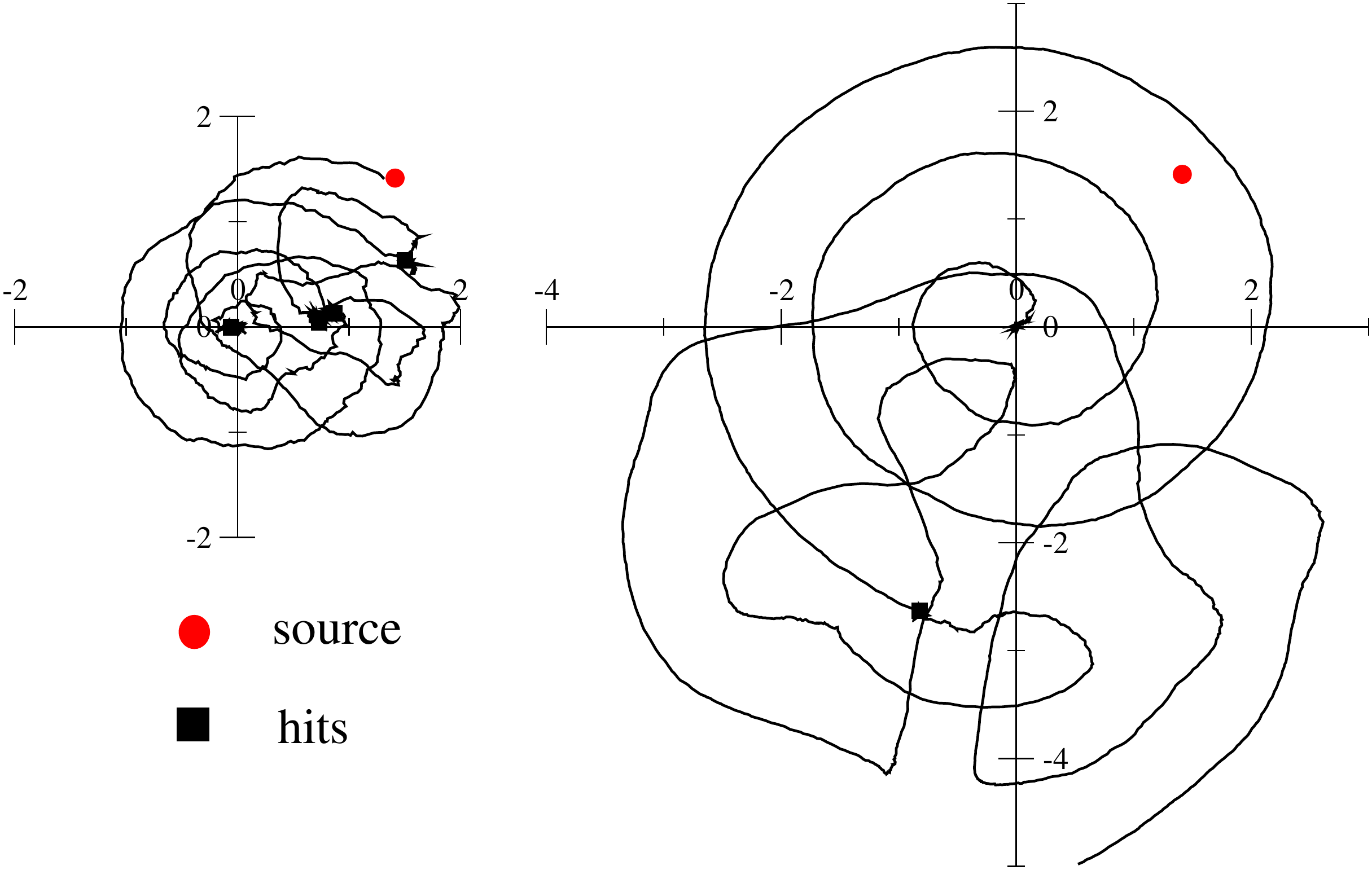}
% traj2d
\includegraphics[width=4cm]{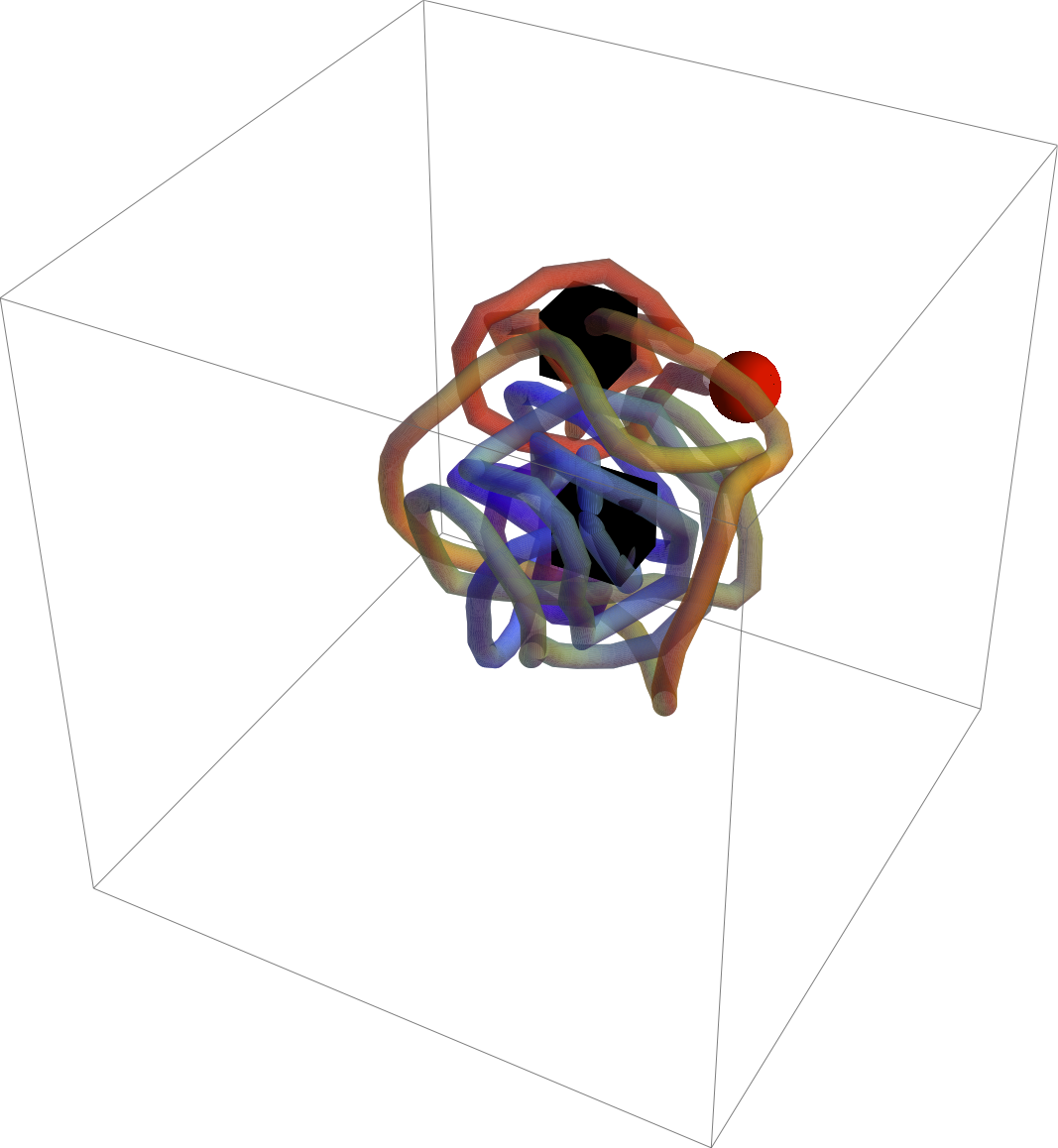}
% 3dnotfound.png
\includegraphics[width=4cm]{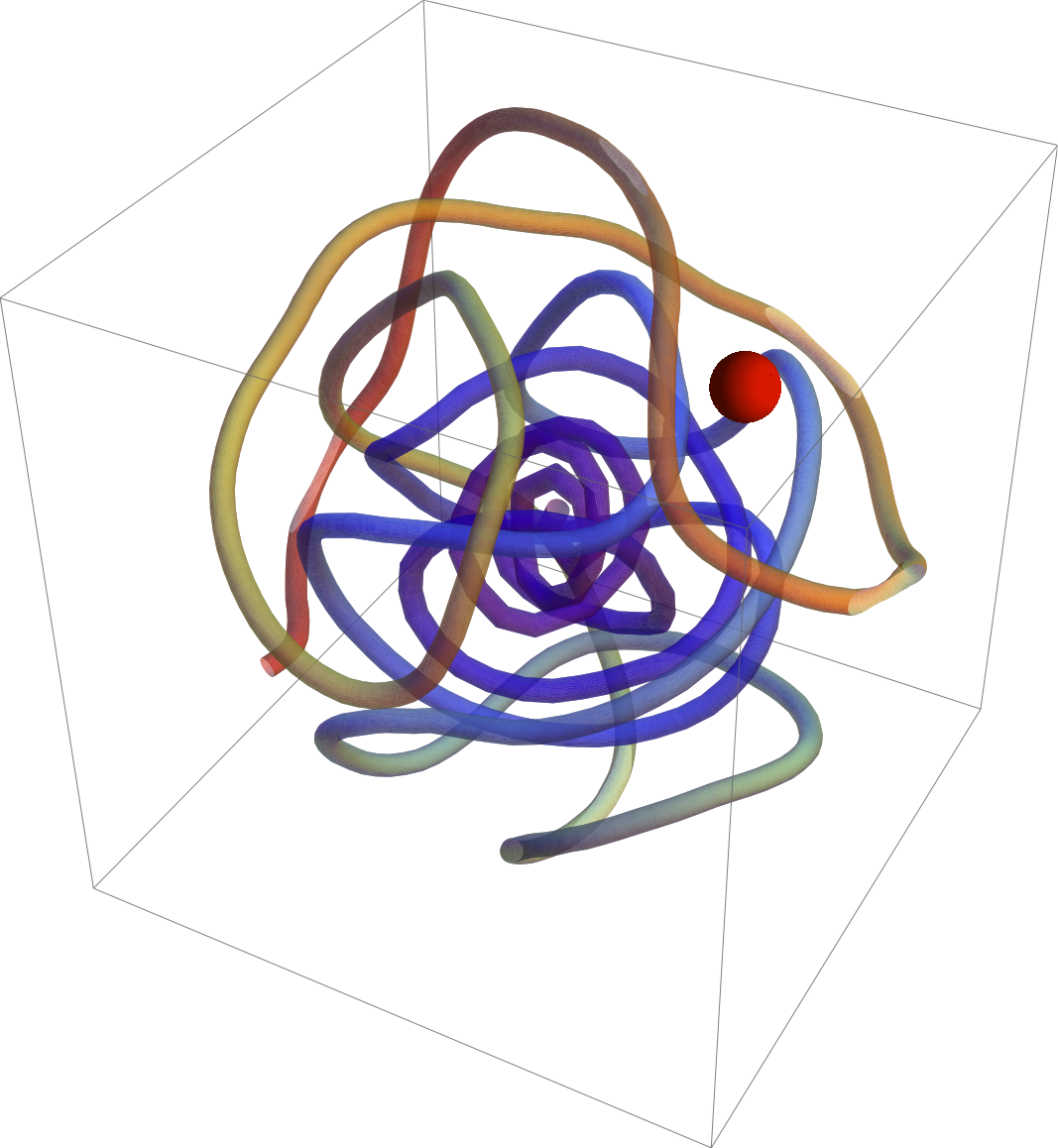}
% 3dfound.png
\vskip -.4cm
\caption{Examples of search trajectories with hits in $D=2$ (top, $\gamma=.02$) and $D=3$ (bottom, $\gamma=.01$) dimensions. Trajectories on the left find the source, while searches on the right are not successful. The initial distance to the source is $d_{0}=2$. Red disks or spheres represent points at distance $<d_{\text{halt}}$ to the source. Black squares or cubes locate the hits; their size correspond to the amplitude of the erratic motion of the searcher during the pinning after a hit.}
\label{fig:trajhit}
\end{figure}

Examples of trajectories where the searcher receives hits at times $t_i>0$ are shown in Fig.~\ref{fig:trajhit}. Generally speaking, the trajectories are denser when more hits are received.  After each hit $i$, the posterior distribution $P_t$ is considerably reinforced in ${\bf x}(t_i)$ and its neighborhood. The searcher remains in the immediate vicinity for a certain time, $t_w$, until the search resumes. Informally speaking, the searcher makes sure that the source is not in ${\bf x}(t_i)$ before looking elsewhere. 
Imagine that the searcher has not moved at all for a period of time $t_w$ after the hit at time $t_i$. The posterior distribution is then 
\begin{equation}
P_{t_i+t_w} ({\bf y}) \propto P_{t_i^-}({\bf y}) \; R({\bf y}-{\bf x}(t_i)) \, e^{-t_w\; R({\bf y}-{\bf x}(t_i))} \ .
\end{equation}
Assuming the posterior distribution right before the hit, $P_{t_i^-}$, is smooth in the vicinity of ${\bf x}(t_i)$, the potential $V_{t_i+t_w} ({\bf x}(t_i) + {\bf u})$ is a function of the small displacement $u=|{\bf u}|$ and $t_{w}$ only. There are a local maximum in $u=0$, since $\alpha_1>0$ in (\ref{eq:eqapprox}), and a global minimum in $u_m (t_w)>0$. For $t_w<.4$, $u_m<.01$ is smaller than the error on the position deriving from our Monte Carlo integration, while for $t_w>.5$, $u_m>.1$, and the displacement of the searcher is easily seen.

The pinning effect is an important feature of the continuous formulation of Infotaxis, and was not observed on a lattice. The reason is that, at each time-step, a whole lattice site probability was set to zero in \cite{VERG07}, which created a strong repulsive effect for the searcher and prevented pinning. In the continuous space, however, the trajectory has zero measure and the repulsion is too weak to override the pinning.  As the searcher gets closer to the source, the average delay $\tau$  between successive hits gets smaller. When $\tau \simeq t_w$, the searcher could, in principle, come to a complete halt. The distance from the source for which this happens, $d_{\text{halt}}$, depends on the dimension: $d_{\text{halt}}=.1$ for $D=2$, $d_{\text{halt}}=.3$ for $D=3$. 

%%%%%%%% performances

\section{Performances}

We now introduce a source and observe the trajectory of the searcher reacting to hits (Fig.~\ref{fig:trajhit}). We are interested in the probability that the search process is successful as a function of the initial distance to the source, $d_{0}$. If the searcher reaches the neighborhood of the source of radius $d_{\text{halt}}$ defined above the source is declared found. If the searcher misses this neighborhood and reaches a distance $d_{\text{fail}}\gg1$ to the source such that new hits are highly unlikely, the searcher is declared to be lost. Examples of successful and unsuccessful trajectories are shown in Fig.~\ref{fig:trajhit}.

\begin{figure}
\begin{center}
\includegraphics[width=8.7cm]{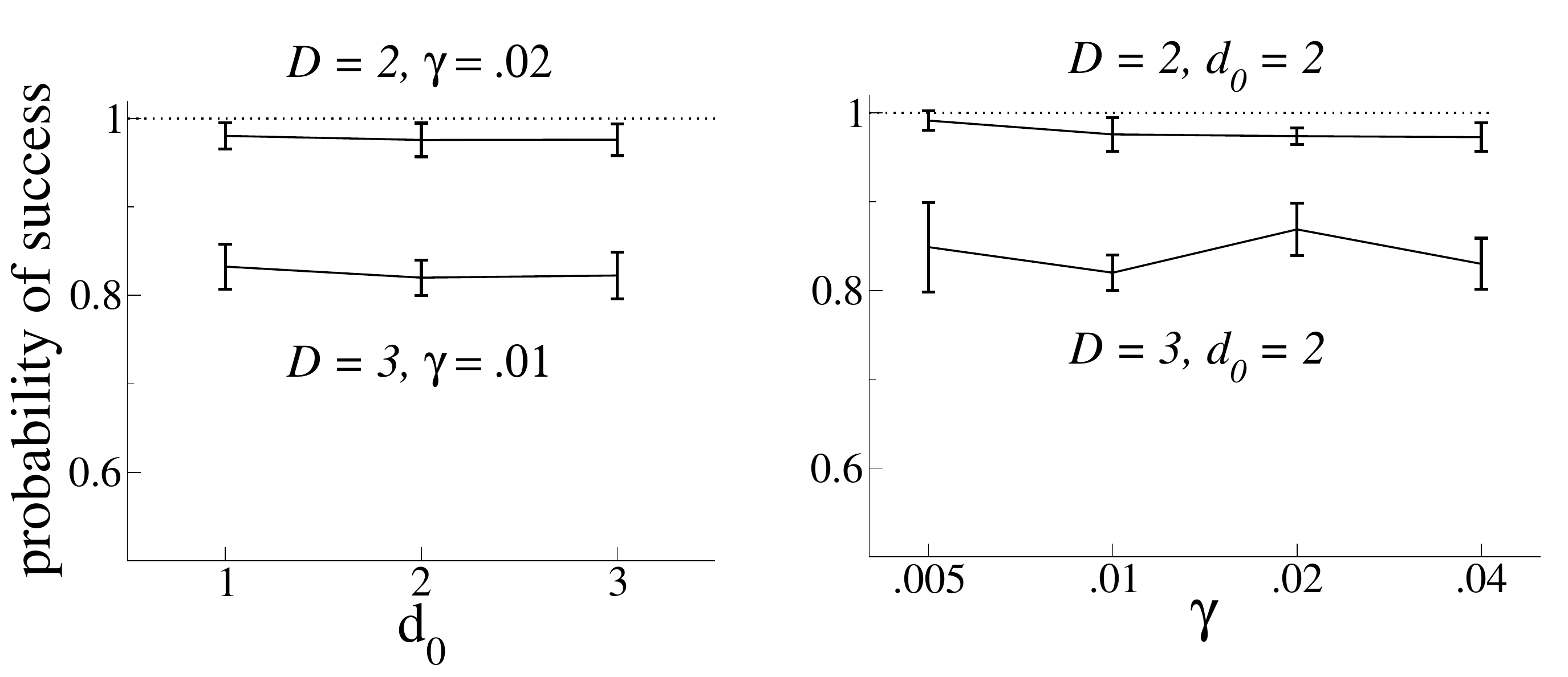}
%fig-proba
\caption{Probability of success of Infotaxis as a function of the initial distance to the source, $d_{0}$ (left), and of the friction $\gamma$ (right). Top points correspond to $D=2$ dimensions, bottom points to $D=3$. The numbers of runs is of about 200 for each point. All probabilities where obtained with $d_{\text{fail}}=8$.}
\label{fig:proba}
\end{center}
\end{figure}

Figure \ref{fig:proba} (left) shows that the probability of success in dimension $D=2$ is compatible with unity for all distances $d_{0}$ smaller than a few units, in agreement with the findings of \cite{VERG07}. On the contrary, in dimension $D=3$, the probability of success is definitely smaller than one, and is about $.8$ for distances $d_0$ ranging from 1 to 3 and for $\gamma=.01$. We have also measured the probability of success at fixed $d_{0}$ over a range of values of $\gamma$ and found little variation (Fig.~\ref{fig:proba}, right).

\begin{figure}
\begin{center}
\includegraphics[width=8.7cm]{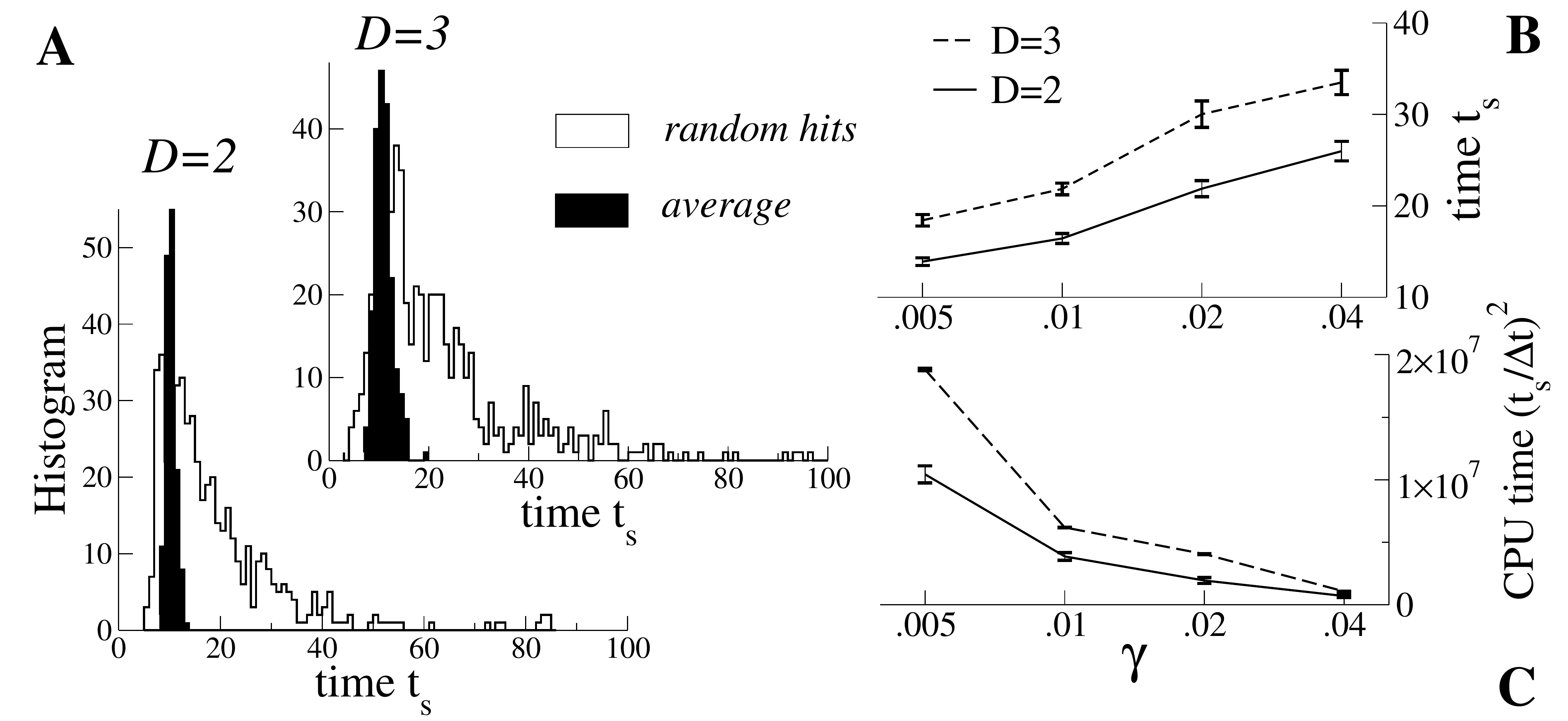}
% fig-histo
\caption{{\bf A.} histograms of the search times $t_{s}$ in $D=2$ ($\gamma=.02$) and $D=3$ ($\gamma=.01$) dimensions for an initial distance $d_{0}=2$ to the source. Full histograms correspond to the average trajectories, contour histograms to trajectories with random hits. {\bf B.} Average search time $t_{s}$ as a function of $\gamma$. {\bf C.} total CPU time as a function of $\gamma$, calculated as $(t_s/\Delta t)^2$. }
\label{fig:histo}
\end{center}
\end{figure}

Figure~\ref{fig:histo}A shows that the distribution of the search times $t_{s}$ of successful runs has a positive skew and a roughly exponential tail not only in dimension $D=2$ \cite{VERG07} but also in dimension $D=3$. The exponential nature of the tail was checked for various values of the distance $d_0$. Figure~\ref{fig:histo}B shows that the average time to find the source, $t_{s}$, decreases with $\gamma$. The CPU time scales as $A\, (t_{s}/\Delta t)^2$, where $\Delta t=\gamma$ is chosen to obtain numerical stability, {\em i.e.} small enough local moves at any step $\ell$. Hence, the CPU time is much larger for small friction (Fig.~\ref{fig:histo}C). We find $A\simeq 3$~ms on one core of a 2.4 GHz Intel Core 2 Quad desktop computer, and for $N_{MC}=10^4$ Monte Carlo steps. The CPU time can be decreased by choosing a smaller value for $N_{MC}$, or by increasing $\gamma$ (and $\Delta t$) with time.

\section{Case of a time-dependent friction $\gamma(t)$}
\begin{figure}
\begin{center}
\includegraphics[width=7.5cm]{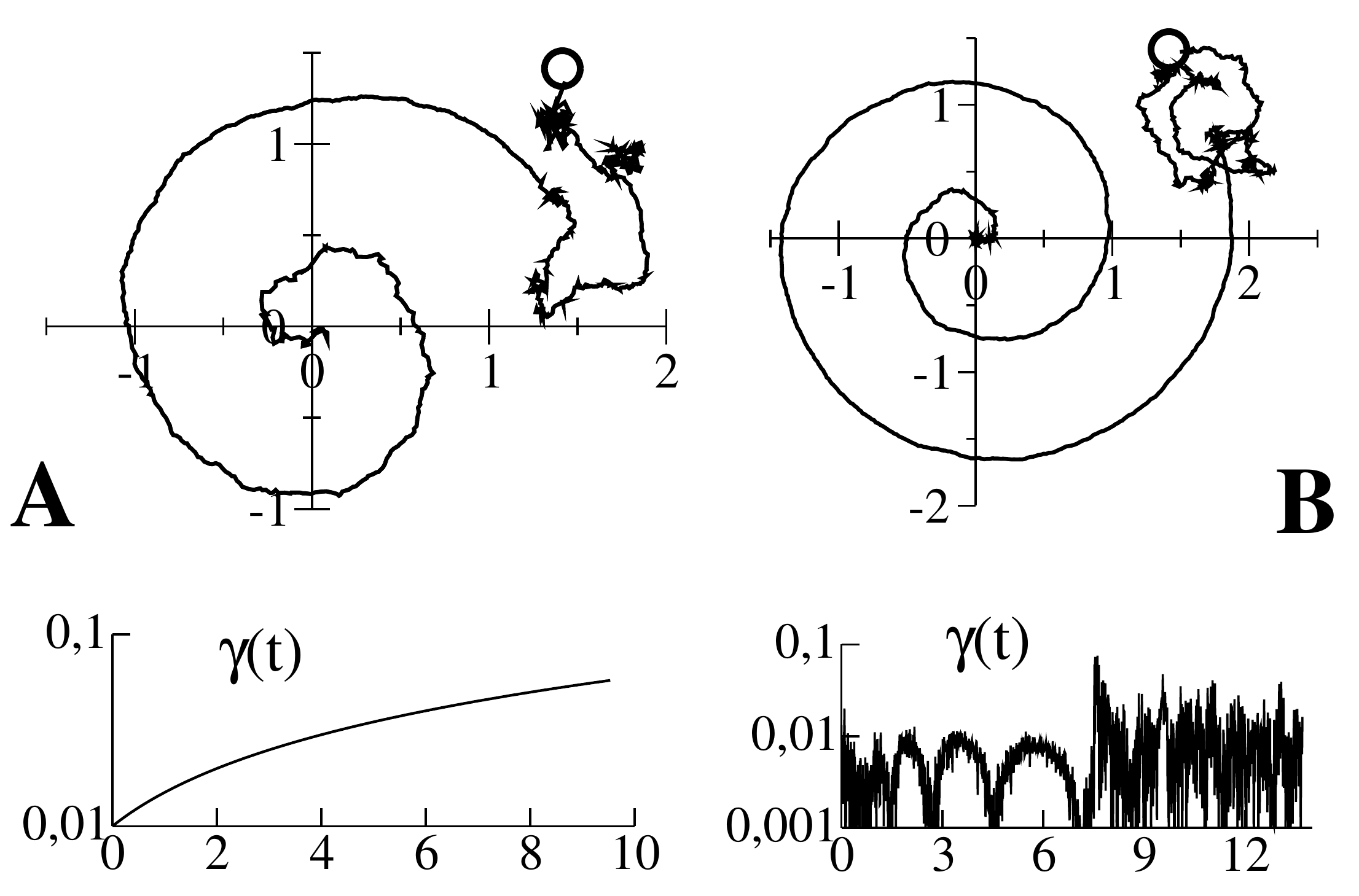}
% fig-fv2
\caption{{\bf A.} A search trajectory for the time-dependent $\gamma (t)$ shown below. {\bf B.}  A trajectory with fixed velocity $v_{0}=2$, and $\gamma (t)=|\nabla_{\bf x}V_{t}|/v_{0}$. The source is located in $(\sqrt 2, \sqrt 2)$ (circle). }
\label{fig:fv}
\end{center}
\end{figure}

The results reported so far correspond to constant frictions $\gamma$. We have generated trajectories with various time-dependent functions $\gamma(t)$, {\em e.g.} slowly increasing to reduce the scaling of the CPU time with $t$ (Fig.~\ref{fig:fv}A), or with the fixed-velocity requirement (Fig.~\ref{fig:fv}B). From a qualitative point of view, we observe no drastic difference with the constant $\gamma$ case, provided that $\gamma(t)$ does not exceed the maximal value $(\simeq .08)$ at which $b\sim 1$ and spirals break down. For instance, the distance between turns in the spiral region of the trajectory in Fig.~\ref{fig:fv}A can be deduced from the value of $b(\gamma(t))$ in Fig.~\ref{fig:spiral2D}C. In the fixed-velocity case of Fig.~\ref{fig:fv}B, before the first hit is detected, $\gamma(t)$ shows regular oscillations and the trajectory has a spiral-like shape with $b$ corresponding to the maximum value of $\gamma(t) (\simeq .011)$. Larger fluctuations are visible during the erratic motion after each hit, but $\gamma(t)$ remains smaller than the arrest value $\gamma\simeq .08$.

We have seen in Fig.~\ref{fig:proba} that the probability of success is essentially independent of $d_{0}$ and $\gamma$. This key result can be understood as follows. Far away from the source, hits are very unlikely and the trajectory develops as a spiral. As the area spanned by the trajectory is roughly independent of $\gamma$ (Fig.~\ref{fig:spiral2D}B), the time it takes for the searcher to reach the $D$--dimensional sphere of radius 1 and centered in the source will be roughly independent of $\gamma$ and will strongly increase with $d_{0}$, while the time spent in the sphere and the number of hits received will be essentially independent of both $\gamma$ and $d_{0}$. We conclude that varying $\gamma$ has little consequence on the performance of Infotaxis, as long as the spiral-like motion is possible. 

\section{Motion in presence of the 'average' signal} 
Certain characteristics of the search time distribution, such as the typical (most probable) value of $t_{s}$, can be assessed from the study of an abstract searcher receiving an average signal rather than discrete and sparse hits. To define an average posterior density $P_{t}^{av}({\bf y})$ we remark that $P_{t}({\bf y})$ contains the product of $N_H$ stochastic $R$ factors over the hits in (\ref{eq:py}). It is natural to define $P_{t}^{av}({\bf y})$ through the average value of $\log P_{t}({\bf y})$ over the hits:
\begin{equation}
P_{t}^{av}({\bf y})\propto e^{-\int_0^t \upd t^\prime R({\bf y}-{\bf x}(t^\prime))+R({\bf y}^*-{\bf x}(t^\prime))\log R({\bf y}-{\bf x}(t^\prime))} P_0({\bf y})
\label{eq:postprob}
\end{equation}
up to a multiplicative normalization constant; here, ${\bf y}^*$ denotes the location of the source as usual. We define the average search trajectory as the solution of equation (\ref{eq:eqsearcher}) with the average $\langle \cdot \rangle$ (\ref{bracket}) calculated over the measure (\ref{eq:postprob}). Average search trajectories are obviously smoother than trajectories with random hits, but have common features, such as a possible return towards the origin after having been close to the source.

Figure \ref{fig:histo}A shows that the search time for the average motion is in very good agreement with the typical search time for trajectories with random hits. This coincidence has been observed for all the frictions $\gamma$ and distance $d_{0}$ to the source we have tested. Note that, while the average motion is fully deterministic, some noise is introduced to break the rotational invariance and select the initial phase of the spiral; this noise is responsible for the small width of the full histograms shown in Fig.~\ref{fig:histo}A. 

\begin{figure}
\begin{center}
\includegraphics[width=8cm]{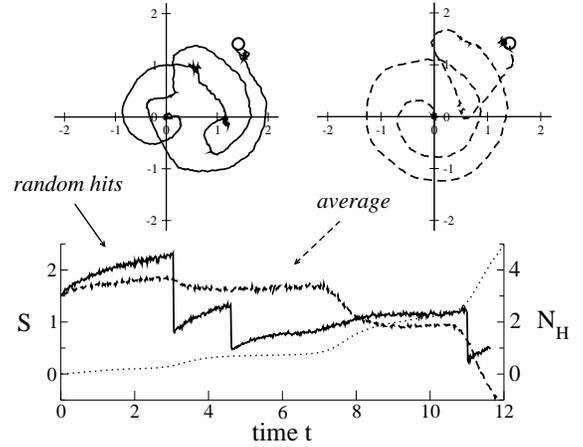}
% entropy
\caption{Entropy $S(t)$ (bottom, left scale) for one trajectory ${\bf x}(t)$ obtained with random hits (top left, full curve, 3 hits are received) and the average trajectory (top right, dashed curve). The dotted line shows the average number of hits $N_{H}$ (right scale) received along the average trajectory. The source is located in $(\sqrt 2,\sqrt 2)$ (circle).}\label{fig:entro}
\end{center}
\end{figure}

Figure~\ref{fig:entro} shows the average search trajectory and a random trajectory in $D=2$ dimensions, together with the entropies of their posterior distributions. In the random case, the entropy abruptly decreases right after a hit, then increases until the next hit is received (due to the exponential time decay in $P_t$ (\ref{eq:py})). As for the average motion, the entropy shows weak oscillations (due to the spiral motion) superimposed to a smooth trend, which decreases as the searcher gets close to the source. 

\section{Conclusion}
In this letter we have presented a continuous-space version of Infotaxis, and have analyzed its behavior in two and three dimensions. When the initial distance to the source, $d_0$, is of the order of the decay length of $R$, the probability that Infotaxis finds the target is essentially equal to unity in $D=2$, and is smaller ($\simeq .8$) in $D=3$ dimensions. The probability of success is roughly independent of $\gamma(t)$ in (\ref{eq:eqsearcher}), while the search time and the CPU time strongly depend on the friction. The quadratic increase of the CPU time and the presence of the pinning effect make the computational cost increase as the searcher gets closer to the source. Note that the CPU time could be made linear in $t_{s}$ (instead of quadratic) if the integral in (\ref{eq:py}) were restricted to the recent past, {\em i.e.} if the search had a finite memory.
  
%Our work could be extended in many directions. We have focused only on the case where the source emits hits without drift in one specific direction, called wind in \cite{VERG07}. While the presence of a drift presumably makes the search process easier, it would be interesting to analyze the shape of the corresponding three-dimensional trajectories. 

The pinning of the trajectory by the hits is a direct consequence of the greedy nature of Infotaxis: the searcher moves to maximize the immediate gain in information, irrespectively of what could be gained on a longer time horizon. To overstep this greedy strategy consider the expected gain in information, $I[{\bf x} (t'); t<t'<t+\tau]$, when the searcher plans to move along a portion of trajectory ${\bf x} (t')$ during the current time $t$ and the time $t+\tau$. The best portion of trajectory is determined through the maximization of $I$, minus a quadratic term $\propto \gamma\, \dot {\bf x}^2$ penalizing large velocities. While this variational calculation appears intractable for general $\tau$, a systematic expansion in powers of $\tau$ is possible. To the lowest orders in $\tau$ the equation of motion becomes
\begin{equation}\label{ng}
\tau^2 \big(\nabla_{\bf x} \nabla_{\bf x} V_t ({\bf x}) \big) \,\ddot {\bf x} + \gamma \,\dot {\bf x} = -  \nabla_{\bf x} V_t ({\bf x})
\end{equation}
up to $O(\tau)$ corrections to the friction $\gamma$ and to the force on the right hand side. The introduction of a finite-time horizon, $\tau$, gives birth to an inertial term, with an effective mass tensor proportional to the curvature matrix of the potential (\ref{eq:pot}). This inertial motion could help reduce the pinning following a hit, and avoid the slowing down of the searcher close to the source. The analysis of (\ref{ng}) and of the search trajectories is left for a future work.

Last of all, the shapes of the trajectories observed in two and three dimensions result from a trade-off between the self-repulsion of the trajectory (the searcher does not come again close a point where the source was not detected) and the confinement due to the hits or to the prior (the source is likely to be close to a detection). This trade-off is present in physical systems such as polyelectrolytes (charged polymers) confined in a volume or on a surface \cite{ANGE}. It is however unclear how far the analogy between the out-of-equilibrium process generated by Infotaxis and equilibrium polyelectrolytes could be pursued \cite{amit}.

\vskip .1 cm\noindent
{\bf Acknowledgment:} we thank J-F. Joanny, S. Leibler, A. Libchaber, T. Maggs, M. Vergassola for useful discussions. This work was partially funded by the ANR 06-JCJC-051 and 09-BLAN-6011 grants.

\end{document}